\documentclass[useAMS,usenatbib]{mn2e}

\usepackage[pdftex]{graphicx}

\title[Quasar orientation indicators]{Comparing different indicators of quasar orientation}
\author[K. Van Gorkom et al.]{Kyle J. Van Gorkom$^{1}$, John F. C. Wardle$^{1}$\thanks{E-mail: wardle@brandeis.edu}, Andreas P. Rauch$^{1}$, and Doug B. Gobeille$^{2}$\\
$^{1}$Department of Physics MS-057, Brandeis University, Waltham, MA 02454-0911\\
$^{2}$Department of Physics, University of South Florida, Tampa, FL 33620-5700}
\begin{document}

\date{draft: \today}

\pagerange{\pageref{firstpage}--\pageref{lastpage}} \pubyear{2015}

\maketitle

\label{firstpage}

\begin{abstract}
Radio core dominance, the rest-frame ratio of core to lobe luminosity, has been widely used as a measure of Doppler boosting of a quasar's radio jets and hence of the inclination of the central engine's spin axis to the line of sight. However, the use of the radio lobe luminosity in the denominator (essentially to try and factor out the intrinsic power of the central engine) has been criticized and other proxies for the intrinsic engine power have been proposed. These include the optical continuum luminosity, and the luminosity of the narrow-line region. Each is plausible, but so far none has been shown to be clearly better than the others. In this paper we evaluate four different measures of core dominance using a new sample of 126 radio loud quasars, carefully selected to be as free as possible of orientation bias, together with high quality VLA images and optical spectra from the SDSS. We find that normalizing the radio core luminosity by the optical continuum luminosity yields a demonstrably superior orientation indicator. In addition, by comparing the equivalent widths of broad emission lines in our orientation-unbiased sample to those of sources in the MOJAVE program, we show that the beamed optical synchrotron emission from the jets is not a significant component of the optical continuum for the sources in our sample. 
We also discuss future applications of these results.

\end{abstract}


\section{Introduction}

The notion of core dominance as an indicator of the orientation of a radio source to the line of sight was first suggested in an important paper by Orr and Browne (1982). This paper was also the first to identify core-dominated quasars as ordinary FR II quasars viewed at a small angle to the parsec-scale jets.

Orr and Browne defined the radio core dominance, $R$, as the ratio of the radio flux density of the quasar core observed at arcsecond resolution to the flux density of the extended lobe emission. Both flux densities are appropriately K-corrected for frequency and redshift to reflect rest-frame values. It is widely accepted that the numerator of $R$ is proportional to the Doppler beaming factor of the pc-scale jets, $\delta^{(2+\alpha)}$, where the $\delta$ is the Doppler factor and $\alpha$ is the spectral index (defined as $S_{\nu} \propto \nu^{-\alpha}$). The Doppler factor is $\delta = (\gamma(1-\beta\cos\theta))^{-1}$ where the jet has speed $\beta c$, Lorentz factor $\gamma$ and makes an angle $\theta$ to the line of sight. We ignore radiation from the backward pointing jet. The beaming factor is a strong function of the angle $\theta$, so the core flux density indeed reflects the orientation of the jets to the line of sight. However, the core flux density must also depend on the power of the central engine, and the unbeamed core luminosity function is broad (Cara \& Lister 2008). The denominator in Orr and Browne's expression is an attempt to factor out this dependence.

While the flux density of the extended emission undoubtedly depends on the time averaged power of the central engine, it depends on other factors as well, such as the gaseous environment and the history of the source; it may not be a good measure of the {\em present} engine power reflected by the core flux density. These points were made by Wills and Brotherton (1995), who suggested that the continuum optical flux density (they used K-corrected V-band magnitudes) might serve as a better measure of the intrinsic engine power. They denoted the ratio of the radio core flux density to the optical continuum flux density (or equivalently, the ratio of the radio core luminosity to the optical continuum luminosity) as $R_V$.

Wills \& Brotherton pointed to supporting evidence for this connection in the results of Yee \& Oke (1978) and Shuder (1981), who found a proportionality between the optical continuum luminosity and emission-line luminosity, and Rawlings \& Saunders (1981), who found a proportionality between the jet power and the luminosity of the narrow-line region.

They also showed that expected correlations with orientation dependent properties tended to be stronger with $R_V$ than with $R$ (see their Table 1). But the use of $R_V$ has certainly not gained universal acceptance as the superior orientation indicator. Since 1996, citations to Orr and Browne (1982) still vastly outnumber those to Wills and Brotherton (1995). More recently some authors have used both $R$ and $R_V$ in their correlations (e.g. Barthel et al. (2000), Richards et al. (2001), Aars et al. (2005), Kimball et al. (2011)). In none of these papers is a strong conclusion drawn in favor of one measure over the other as an indicator of orientation.

Certainly, both $R$ and $R_V$ are qualitative indicators of orientation to the line of sight because both contain the Doppler beamed radio core luminosity in the numerator. Which core dominance measure is superior then depends on which uses the better proxy for intrinsic engine power in the denominator. That is what we investigate in this paper.

The layout of the rest of the paper is as follows. In section~\ref{sect:LSS} we describe the radio sample that we will use for most of the analysis. Its most important attribute is that it is constructed so as to be as free as possible of orientation bias. In section~\ref{sect:beaming} we use observations of MOJAVE blazars to estimate  to what extent the optical continuum flux is affected by beamed optical radiation from the jets. In section~\ref{sect:4Rs} we introduce two additional possible estimators of intrinsic engine power, and their  corresponding core dominance measures. In sections~\ref{sect:corr} and \ref{sect:testing} we compare and evaluate the four measures of core dominance. In section~\ref{sect:HRsample} we apply the same analysis to the Hough-Readhead (1989) sample of lobe-dominated 3C quasars. In section~\ref{sect:conclude} we summarize our results.


\section{The Lobe Selected Radio Sample}
\label{sect:LSS}

The radio sample analyzed in this paper was constructed by Gobeille (2011) and consists of two parts. The first part is a high redshift sample of 123 radio-loud quasars with redshifts in the range $2.5 \leq z \leq 5.28$ that form a complete flux limited sample ($\geq 70$ mJy at 1.4 GHz). The area of sky covered by this sample stretches from 7 to 17.5 hours in right ascension and from $0^{\circ}$ to $65^{\circ}$ in declination. This area is covered by the Sloan Digital Sky Survey (Abazajian et al. 2003) so excellent optical data are available for each quasar. High resolution radio images made with the Karl G. Jansky Very Large Array\footnote {The National Radio Astronomy Observatory is a facility of the National Science Foundation operated under cooperative agreement by Associated Universities, Inc.} are presented by Gobeille, Wardle \& Cheung (2014). These are a combination of images from archival data for 43 sources and new A-array observations at 1.4 and 5 GHz for the remainder. 

The second part is a low redshift ($z < 2.5$) comparison sample of quasars from the same area of sky, with the same flux limit, and for which high resolution data were found in the VLA archive. We found 131 such objects and re-imaged all of them (Gobeille 2011). They too have optical data from the SDSS.  They range from some of the 3CR quasars deeply imaged by Bridle et al. (1994) to sources with only brief observations. The vast majority of the LSS consists of 4C quasars and some Bologna quasars that were observed with the VLA as soon as optical identifications were made. Most of the observations are to be found in Ulvestad et al. (1981), Hintzen, Ulvestad \& Owen (1983), Owen \& Puschell (1984), Gower \& Hutchings (1984), Barthel et al. (1988), Price et al. (1993), Lonsdale, Barthel \& Miley (1993).
 
A critical point is that in no case was a source observed with any prior knowledge of its structure. Also, in this paper we are not counting radio sources or constructing, say, a luminosity function, so incompleteness is not necessarily an issue provided that we have a \textit{representative} sample of sources. There is no indication that this is not the case.

Every quasar in the combined sample has a clearly visible compact radio component that is coincident with the optical position measured by SDSS. These cores are widely believed to be Doppler boosted pc-scale jets (e.g. Orr \& Browne 1982, Blandford \& K\"onigl 1979), and their contribution to the total flux density of the source is therefore strongly dependent on their inclination to the line of sight. Also, many sources exhibit prominent kpc-scale jets, which we know are also significantly Doppler boosted because they are (in quasars) visible only on one side of the source and this is invariably the side of the source pointing towards us (Garrington et al. 1988). From deep VLA images, their typical velocities are found to be in the range $0.7c - 0.9c$ where $c$ is the speed of light (Wardle \& Aaron 1995). The more extended lobe emission is not significantly beamed, with an upper limit on the average hotspot advance speed of 0.1c (Scheuer 1995).

If we wish to construct a sample of radio sources that is as free as possible of orientation bias, then we must ignore the emission from the cores and the jets, and include only those sources whose extended emission (mainly the lobes) by itself exceeds 70 mJy at 1.4 GHz. This was determined as follows.

In order to separate the core, jet and extended emission flux densities, each source was treated individually depending on its angular size and on the data sets available in the NRAO archives. 

Over 90\% of the sources in the LSS have archival observations available in several arrays and at several frequencies. Where there are both A and B array observations at the same frequency, they have been combined, following the procedure described in Murphy, Browne and Perley (1993) that allows for variation of the core flux density between observations.

In general, the lobe flux density was calculated as the total flux density minus the core flux density minus the jet flux density. These flux densities were determined in DIFMAP (Shepherd 1997). For the total flux density we placed a box around the whole source and added up all the emission. To guard against missing flux due to resolution effects, this was done using lower resolution observations where available.  The total fluxes measured in the image plane were also checked for consistency with the fringe amplitudes on the shortest baselines.

The cores and jets were also measured in the image plane, by model fitting the features in DIFMAP. Often it was necessary to use higher frequency images to separate clearly the jets and cores from the extended emission.  In particular,  for sources smaller than about 10 arcsec, it was better to use C band or even X band images. To refer fluxes back to 1.4 GHz, we assumed spectral indices of 0.0 for the cores and 0.6 for the jets (Bridle et al 1994).

From our combined high and low redshift samples, 126 quasars had lobe emission that by itelf exceeded 70 mJy at 1.4 GHz, and they form what we call the ``lobe selected sample'' or LSS. This sample is similar to the ``lobe dominated sample'' of 3CR quasars defined by Hough and Readhead (1989) and we have six members in common. But our sample is much larger and extends to much higher redshifts. It should be distributed uniformly in $\cos\theta$ up to a cut-off angle of $45^\circ$ that is believed to separate FR II quasars from FR II radio galaxies (Barthel 1989; Wardle \& Aaron 1997).

The radio data measured from our images and used in this paper consist of radio core flux density at 5~GHz, the lobe flux density at 5~GHz (again, computed as total flux minus core flux minus jet flux), the projected arm length (defined as the angular distance from the core to the brighter hotspot), and the bending angle (defined as the external angle between two lines drawn from either hotspot to the core). These data are listed in Rauch et al. (in preparation).


\section{Blazars and Beaming}
\label{sect:beaming}
\begin{figure}
  \includegraphics[width=\linewidth]{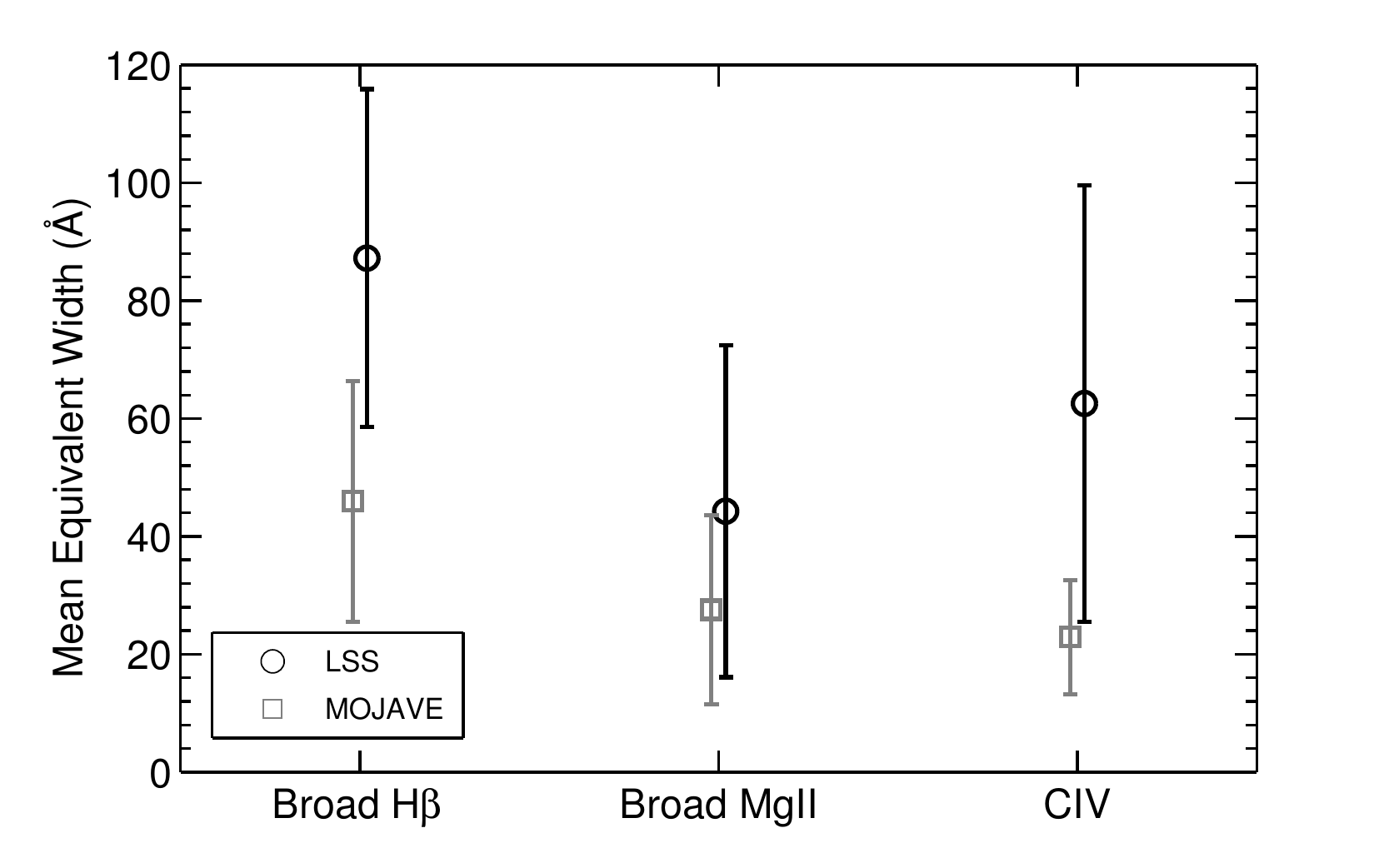}
  \caption{The mean equivalent widths of the broad lines H$\beta$, MgII, and CIV, plotted with sample standard deviations for the LSS sample (circles) and the MM highly beamed quasars (squares). The MM sources show smaller EWs across all lines, consistent with the presence of beaming.}
  \label{fig:ew}
\end{figure}

If the optical continuum luminosity is to function as a proxy for the intrinsic engine power, it cannot show signs of significant beaming. To quantify the amount of optical beaming present in the LSS, we compare it to a sample of blazars. The optical flux of a blazar is polarized and variable and is therefore known to include a beamed synchrotron component. Our blazar sample is drawn from the MOJAVE-Mets\"{a}hovi (henceforth MM) sample of 62 blazars (Savolainen et al. 2010). We exclude sources classified as BL Lac objects leaving a sample of 48 highly beamed quasars. All of them have prominent cores and the vast majority exhibit superluminal motion (Lister et al. 2009). For all 48 sources, long-term total flux density monitoring at 22 and 37 GHz at the Mets\"ahovi Radio Observatory allowed Doppler factors to be calculated from the flux variability (Hovatta et al. 2009). Using optical data for this sample from Torrealba et al. (2012), we have rest-frame equivalent widths (EW) for the broad lines of H$\beta$ (11 quasars), MgII (24 quasars), and CIV (14 quasars). In Figure \ref{fig:ew} we show the mean rest-frame equivalent widths for these three broad lines for the LSS and the MM sources. As expected, the highly beamed quasars show equivalent widths that are systematically smaller than those of the LSS sources, consistent with the presence of beaming in the optical continua of the MM sources. The ratios are listed in Table~\ref{table:beamed}

\begin{table}
\centering
\caption{Ratio of LSS to MM broad line equivalent widths, with standard deviations calculated from those of the sample means}
\label{table:beamed}
\begin{tabular}{lc}
\hline
Emission Line & $\langle \mathrm{EW} \rangle_ \mathrm{LSS} \big/ \langle  \mathrm{EW} \rangle_ \mathrm{MM}$ \\
\hline
H$\beta$&$1.9 \pm 0.3$\\
MgII&$1.6 \pm 0.2$\\
CIV&$2.7 \pm 0.4$\\
Weighted Mean&$1.9 \pm 0.2$\\
\hline
\end{tabular}
\end{table}

We expect the mean MM source to be orientated at a small angle to the line of sight. Thus, the luminosity of the optical continuum will arise from some combination of  thermal radiation from the accretion disk and beamed synchrotron radiation from the jet: 
$L_{\nu, \mathrm{continuum}} = L_{\nu, \mathrm{thermal}} + L_{\nu, \mathrm{beamed}}(\theta)$. The mean LSS source, however, will be oriented at a much larger angle to the line of sight and the beamed contribution to the optical continuum will be negligible: $L_{\nu, \mathrm{continuum}} \approx L_{\nu, \mathrm{thermal}}$. 

We write the equivalent width as $\mathrm{EW} = L_\mathrm{line}^\mathrm{tot} / L_{\nu, \mathrm{continuum}}$ where $L_\mathrm{line}^\mathrm{tot} = \int\limits_\mathrm{line} L_{\nu, \mathrm{line}} \mathrm{d}\nu$. The fact that the average equivalent width for the MM sources is about half that for the LSS sources implies that in the blazar sample $L_{\nu,\mathrm{beamed}}(\theta) \approx L_{\nu,\mathrm{thermal}}$. But $L_{\nu,\mathrm{beamed}}(\theta) \propto \delta^{2+\alpha}(\theta)$, where $\delta$ is the Doppler factor and $\alpha$ is the spectral index. The median Doppler factor for the MM sources is 15.5. If the median source is oriented near the angle that maximizes superluminal motion, given by $\cos\theta = \beta$, then $\theta \approx 3\fdg 7$. A straightforward calculation finds that at angles greater than $7\fdg 1$, $L_{\nu,\mathrm{beamed}}$ drops below $0.1 L_{\nu,\mathrm{thermal}}$ and can be ignored. The fraction of sources oriented between $0^{\circ}$ and $7\fdg 1$ in a sample that is unbiased in orientation (i.e. distributed uniformly in $\cos\theta$) up to the $45^\circ$ cut-off, is $2.6\%$. In our LSS sample of 126 sources, we expect about three sources to have $F_{\nu,\mathrm{beamed}} > 0.1 L_{\nu,\mathrm{thermal}}$. These would presumably be among the more core-dominated sources in the sample, but clearly beamed optical emission is not a major concern for our lobe-selected sample.

In order to directly compare the mean equivalent widths of the LSS and MM samples, we should make sure that the results are not biased by the Baldwin effect (Baldwin 1977), in which the equivalent widths of broad emission-lines vary roughly as $L^{-1/3}_{\nu, \mathrm{continuum}}$. We have checked the values of $L_{\nu,\mathrm{2500 \AA\ }}$ for the sources in both samples and find that they cover very similar ranges. We conclude that the Baldwin effect (which is quite apparent in MgII, for which we have the most data) should affect both samples to a similar extent, and that the differences between the samples evident in Table~\ref{table:beamed} and in Figure \ref{fig:ew} can indeed be attributed to beaming.

It is, of course, not a new result that part of the optical continuum is beamed in some classes of quasar. For instance, Wills \& Browne (1986) found that the [OIII] and H$\beta$ equivalent widths were inversely correlated with R, the radio core dominance. Browne \& Murphy (1987) found that quasars with flat radio spectra have smaller  emission-line equivalent widths than do steep-spectrum quasars. Wills et al. (1992) found a highly significant correlation between the fractional optical polarization and the radio core dominance, R. All of these results are consistent with the simple picture discussed above. What we have done here is attempt to quantify the  effect of beaming, and show that it is not important in a sample that has been selected to be largely free of orientation bias.

\section{Four Measures of Core Dominance}
\label{sect:4Rs}

So far we have considered two definitions of core dominance, or equivalently, two possible proxies for the intrinsic power of the central engine. 
 We can also use the Rawlings \& Saunders (1981) result that there is a proportionality between the jet power and the luminosity of the narrow-line region; i.e. we can construct a third possible core dominance measure that normalizes the core luminosity directly by the total luminosity of the narrow-line region. Following the prescription of Rawlings \& Saunders, we take this to be $L_\mathrm{NLR} = 3  (3  L_\mathrm{[OII]} + 1.5  L_\mathrm{[OIII]})$. Where only one line is available, the other is calculated by the relationship $L_\mathrm{[OIII]} = 4L_\mathrm{[OII]}$. 

Finally, Willott et al. (1999) relate the engine power to the 151~MHz luminosity by the formula $Q_\mathrm{Ox} \approx 3 \times 10^{38}L_{151}^{6/7}\ \mathrm{W}$, where $L_{151}$ is given in units of $10^{28}\ \mathrm{W\ Hz^{-1}\ sr^{-1}}$. Punsley (2005) derived a similar expression based on different physical assumptions, and referred to the expression derived by Willott et al. as ``the Oxford formula." We will do likewise, hence $Q_\mathrm{Ox}$. Given the steep spectra of radio lobes and the flat spectra of core components, the total luminosity at 151 MHz essentially measures the lobe luminosity. It has the possible advantage that it includes diffuse emission that is over-resolved by the VLA. Willott et al. make various approximations in deriving their expression. For instance, they assign all sources the same orientation and the same axial ratio. However it did not seem profitable to pursue more detailed models because they also, of necessity, assumed the same gaseous environment for each source (about which little is known), and the same parameters that enter the equipartition calculation (ditto). These uncertainties will add source to source scatter to the calculation but should not change significantly its functional form.

Here we will consider four different measures of core dominance:
\begin{eqnarray}
R = \frac{L_{\nu,\mathrm{5GHz\ Core}}}{L_{\nu,\mathrm{5GHz\ Lobe}}} \\
R_V = \frac{L_{\nu,\mathrm{5GHz\ Core}}}{L_{\nu,\mathrm{2500 \AA\ }}} \\
R_\mathrm{NLR} = \frac{L_{\nu,\mathrm{5GHz\ Core}}}{L_\mathrm{NLR}} \\
R_\mathrm{Ox} = \frac{L_{\nu,\mathrm{5GHz\ Core}}}{Q_\mathrm{Ox}}
\end{eqnarray}

Luminosities were computed from the observed flux densities using the concordance cosmology (Spergel et al. 2003), and K-corrected to their rest-frame values. The 5~GHz core and lobe flux densities were available for all 126 sources of the LSS, from our VLA images. The 2500 \AA\ optical continuum luminosity was taken from the SDSS quasar catalog (Shen et al. 2011) and was available for 99 sources. (We have not corrected the optical luminosities for reddening, though this is not an insignificant effect. The fraction of radio-loud quasars that are significantly reddened is a function of redshift and faintness, and varies from a few percent to at least 17\% (Glickman et al. 2013)). The [OII] and [OIII] line luminosities for the calculation of $L_\mathrm{NLR}$ were also taken from the SDSS and were available for 66 sources.  The 151~MHz flux densities for the calculation of $Q_\mathrm{Ox}$ were taken from the 6C and 7C surveys (Hales et al. 1990 and Waldram et al. 1996) and were available for 37 sources.

 The Spearman rank correlation test shows that each of the four core dominance measures is highly correlated with all the others. This is expected because they all have the same numerator. The quality of each as a measure of orientation will depend on how well the denominator acts to factor out the intrinsic engine power. We attempt to elucidate this in the next sections.


\section{Correlations and Geometrical Rank}
\label{sect:corr}

\begin{table}
\centering
\caption{Spearman  rank correlations among power proxies}
\label{table:powers}
\begin{tabular}{lccccc}
\hline
Parameter& $L_\mathrm{\nu, 5GHz\ Lobe}$ & $L_\mathrm{\nu, 2500\AA}$ & $Q_\mathrm{Ox}$ & $L_\mathrm{NLR}$\\
\hline
$L_\mathrm{\nu, 5GHz\ Lobe}$&1&\ldots&\ldots&\ldots\\
&0&&&\\
$L_\mathrm{\nu, 2500\AA}$&0.55&1&\ldots&\ldots\\
&$<10^{-8}$&0&&\\
$Q_\mathrm{Ox}$&0.76&0.54&1&\ldots\\
&$<10^{-7}$&0.0018&0&\\
$L_\mathrm{NLR}$&0.39&0.41&0.19&1\\
&$<10^{-3}$&$<10^{-3}$&0.46&0\\
\hline
\end{tabular}

\medskip
The Spearman coefficient $\rho$ is given above the two-tailed probability of the correlation arising by chance. 
\end{table}

\begin{figure}
\includegraphics[width=\linewidth]{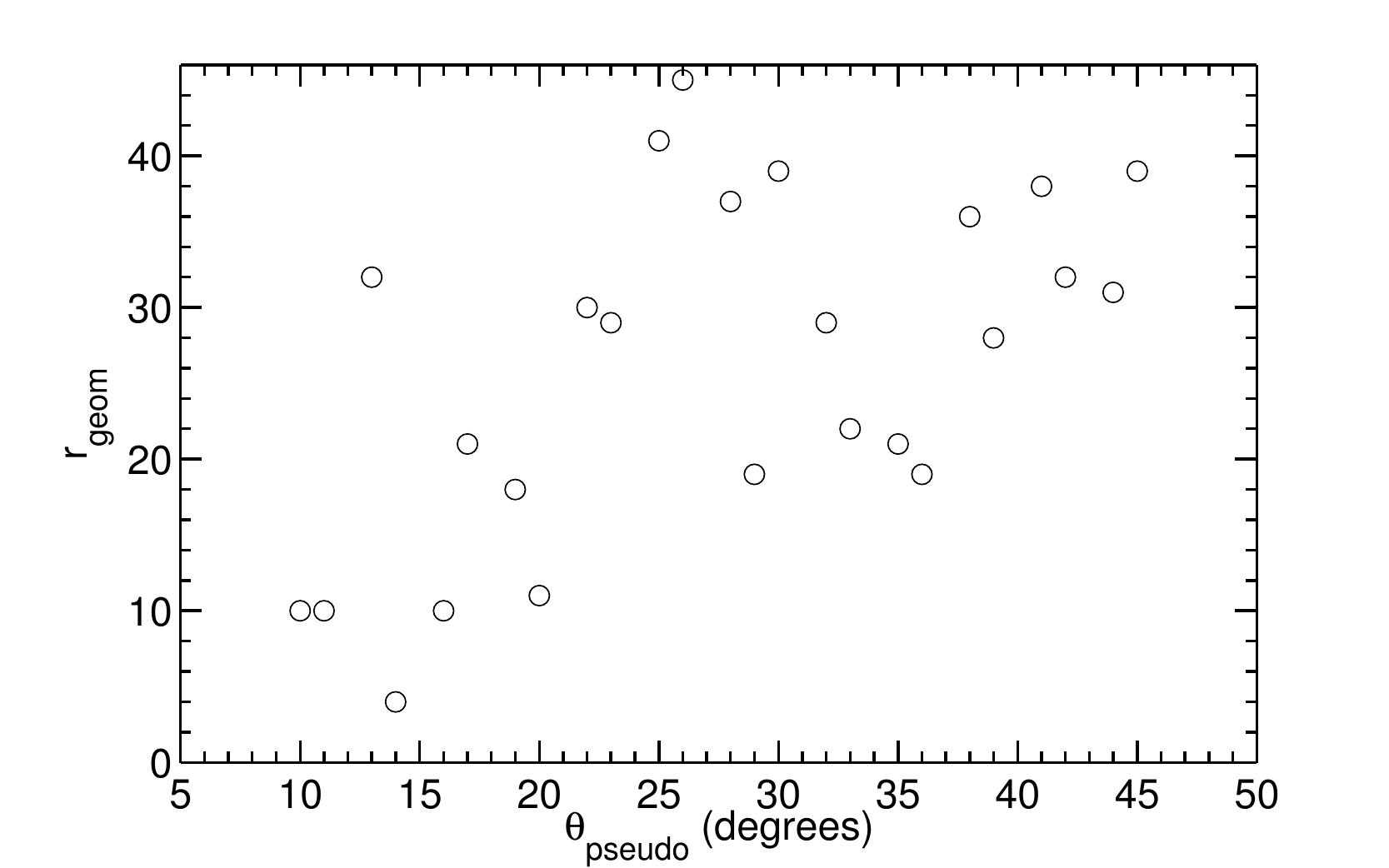}
\caption{The correlation between geometric rank $r_\mathrm{geom}$ and pseudo-angle $\theta_\mathrm{pseudo}$ for the LDQ sample ($\rho$ = 0.53, p-val = 0.0062).}
\label{fig:pseudo}
\end{figure}

\begin{table}
\caption{Spearman rank correlations with geometric rank $r_\mathrm{geom}$}
\label{table:rgeom}
\begin{tabular}{l ccc}
\hline
& Spearman $\rho$ & Two-tailed p-value & variance\\
\hline
$R$ & -0.27  &  0.0021  & 0.81 \\
$R_V$ &   -0.31 &  0.0018 & 0.51 \\
$R_\mathrm{Ox}$ &  -0.33  &  0.043 & 0.73 \\
$R_\mathrm{NLR}$ & -0.25  &  0.045 &  0.87 \\
\hline
\end{tabular}
\end{table}

We list the correlations among the four proxies for engine power in Table \ref{table:powers}. The correlations are significant in all but one case: $L_\mathrm{NLR}$ is not significantly correlated with $Q_\mathrm{Ox}$ at least in this data set ($\rho = 0.19,\ p-value = 0.46$). This suggests that all four proxies do depend on engine power, but to differing degrees, and they are affected by other factors too. All of them will therefore add scatter when used to normalize the radio core luminosity. The ``best'' proxy for the intrinsic engine power should be the one that adds the least scatter.

We now turn to geometrical arguments to help choose the best proxy. Although we have no {\em a priori} knowledge of the orientation of any radio source, information from the radio images can significantly constrain their orientations. For example, sources with the largest projected arm lengths (projected distance from the core to the brighter hotspot) must be oriented near the cut-off angle ($45^{\circ}$). A source with a small arm length, however, could be either intrinsically large and oriented at a small angle or intrinsically small and at a larger angle, so its orientation is not constrained by this argument. Projected size therefore serves as a useful orientation constraint for large sources, less so for medium size sources, and not at all for small sources. (Our results are not very sensitive to the precise value of the cut-off angle, so we do not consider here the ``receding torus'' model, where the cut-off angle depends on luminosity (e.g. Hill et al. 1995; Arshakian 2005).)

Another geometric proxy for orientation is the bending angle of a source. The bending angle is here defined as the exterior angle produced at the intersection of two lines drawn from either hotspot to the core. Any misalignment between the two jets will be exaggerated in a source that makes a small angle to the line of sight. Analysis of the observed bending angles in the LSS shows that they are consistent with modest intrinsic bends drawn randomly from a flat distribution between $0^{\circ}$ and $20^{\circ}$ (Rauch 2013). Thus, sources with large observed bending angles $\gg 20^{\circ}$ must be oriented at small angles to the line of sight. A source that appears straight, on the other hand, may be either intrinsically straight or bent in the plane containing the jet and the line of sight, and its orientation is not constrained.

While both geometric arguments are of limited utility by themselves as orientation indicators, combining them into a single measure usefully begins to sort the sources in the LSS by their orientation. We therefore combine the information in each of these geometrical measures by a summed rank, $r_\mathrm{geom} = r_{l_B} + r_{BA}$. Here, $l_B$ is the projected linear separation between the core and brighter hotspot. Following Rauch (2013), we remove the observed dependence of source size on redshift by the empirically determined expression $l_B = l_{B_{observed}} (1+z)$. We assign the rank $r_{l_B}$ to each source sorted on ascending $l_{B}$ and the rank $r_{BA}$ on descending bending angle. The geometric rank $r_\mathrm{geom} = r_{l_B} + r_{BA}$ should therefore tend to sort the sources by orientation angle, to a considerable extent at large and small ranks, and less so for mid-ranks.

\begin{figure*}
\begin{minipage}{0.4\linewidth}
  \includegraphics[width=\linewidth]{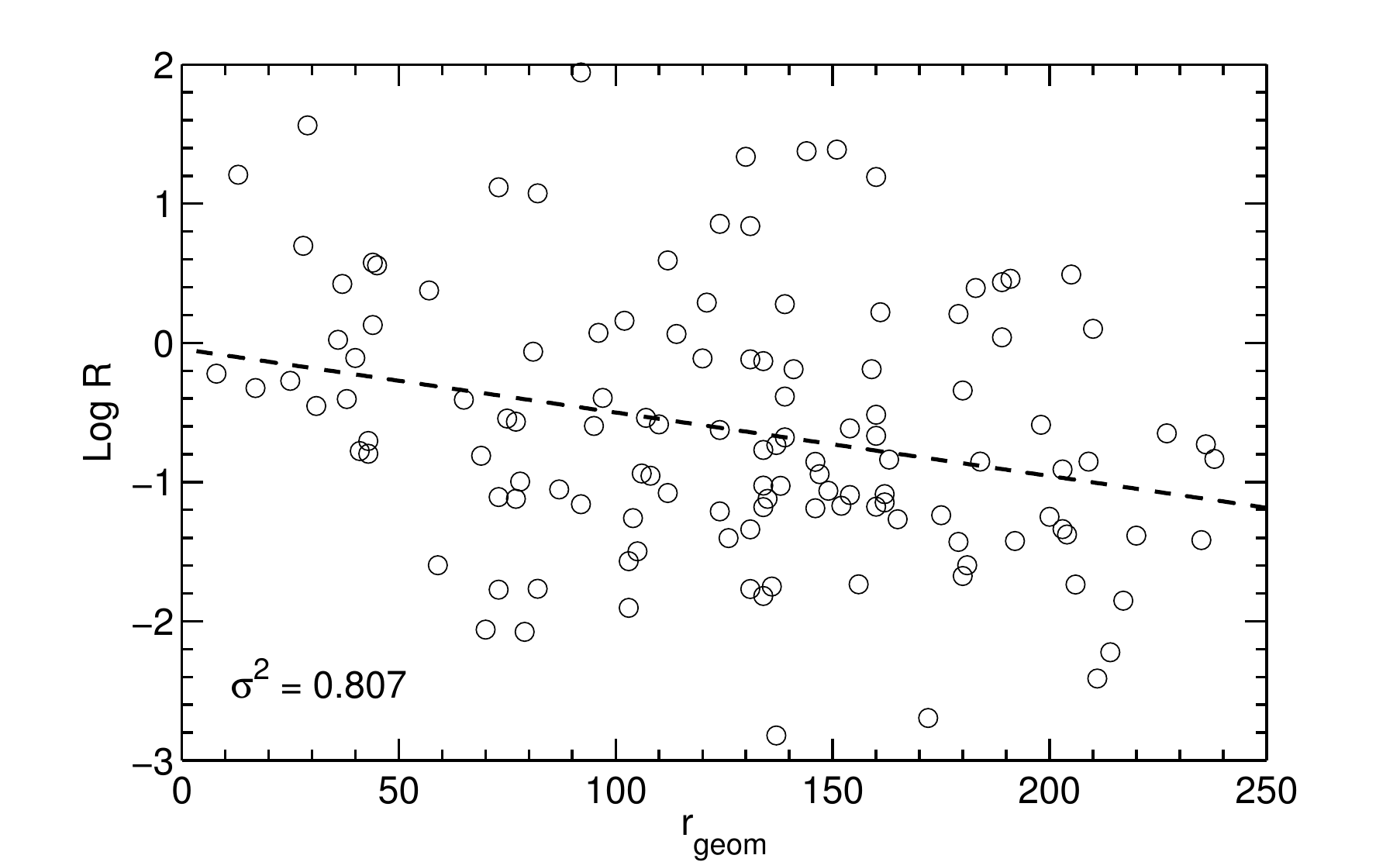}
\end{minipage}
\begin{minipage}{0.4\linewidth}
  \includegraphics[width=\linewidth]{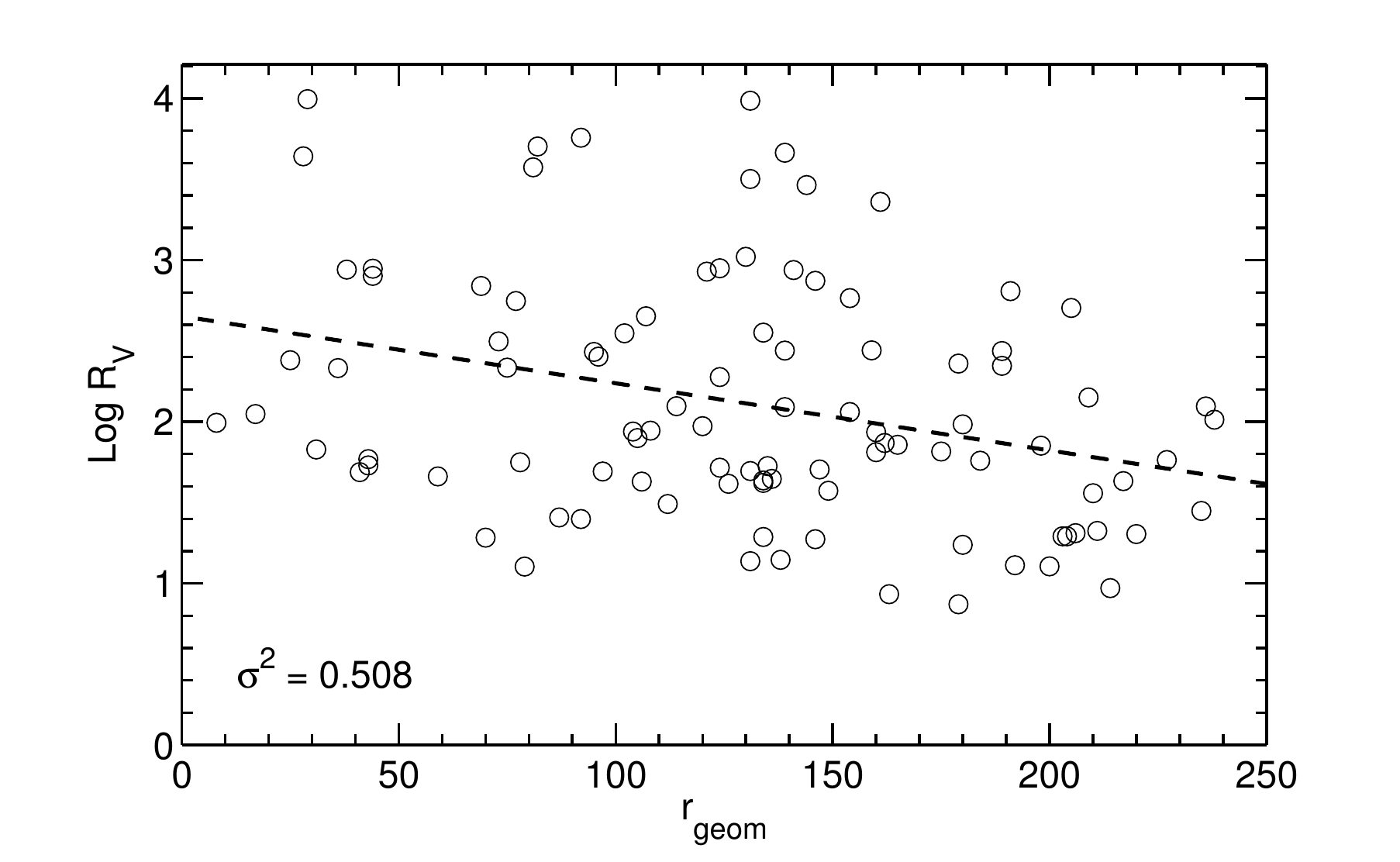}
\end{minipage}
\begin{minipage}{0.4\linewidth}
  \includegraphics[width=\linewidth]{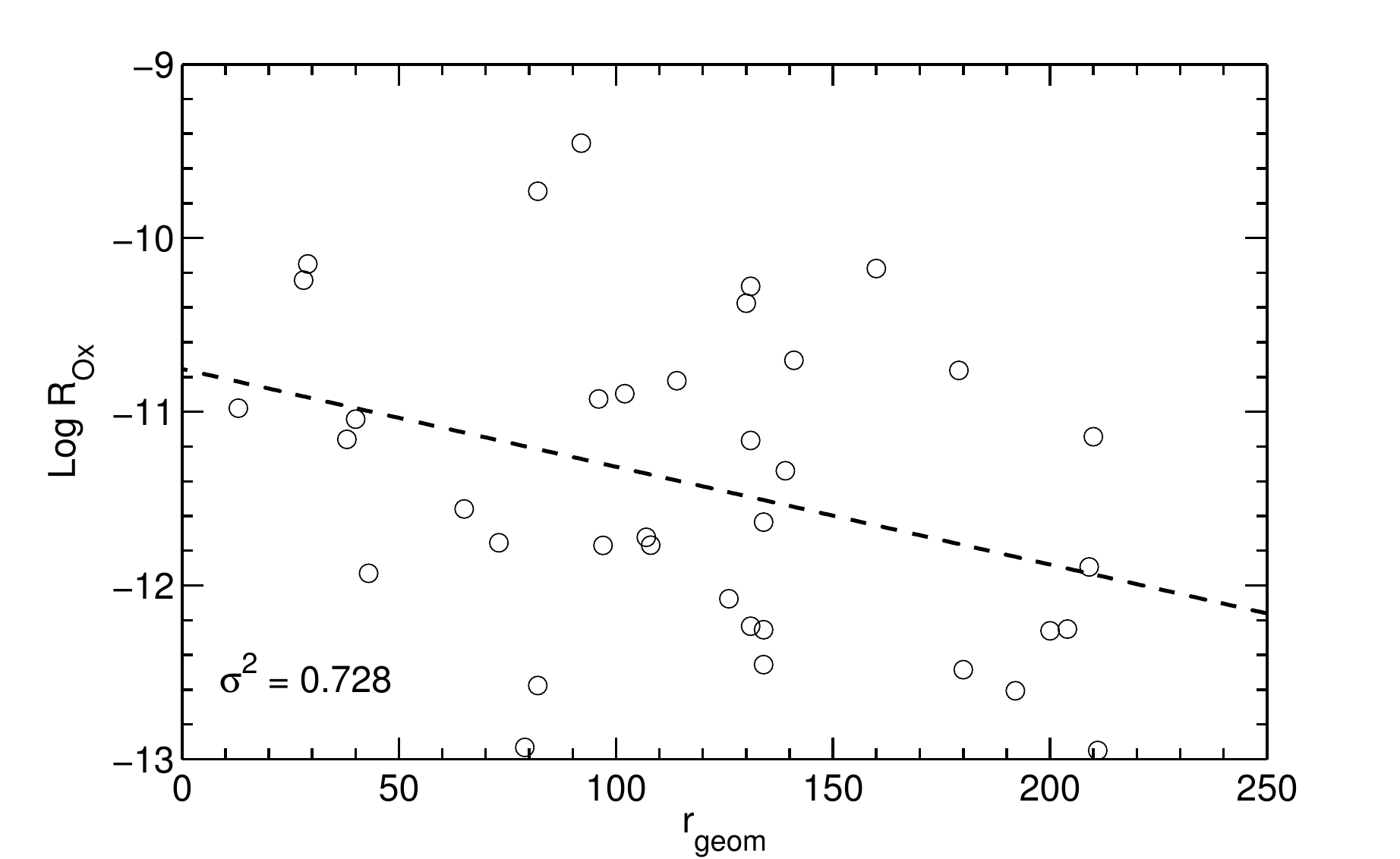}
\end{minipage}
\begin{minipage}{0.4\linewidth}
  \includegraphics[width=\linewidth]{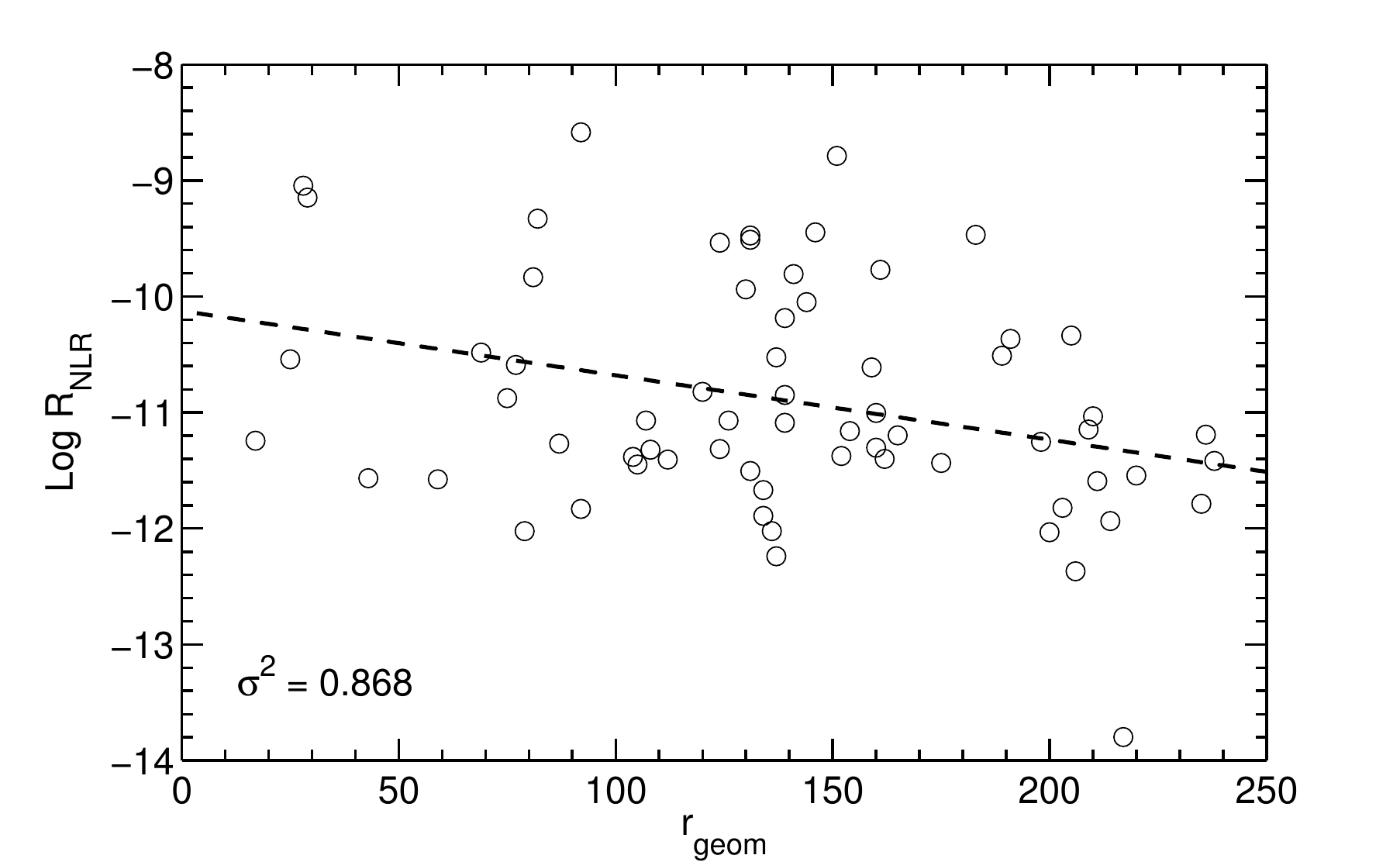}
\end{minipage}

\caption{Core dominance parameters plotted against geometric rank $r_\mathrm{geom}$. The variance $\sigma^2$ about the best fit line is given in the lower left of each plot.}
\label{fig:rgeom}
\end{figure*}

Aars et al. (2005) defined a similar orientation measure for the Hough-Readhead (1989) sample of lobe-dominated  3CR quasars (see section~\ref{sect:HRsample}) that they called $\theta_\mathrm{pseudo}$. They constructed it from a rank ordering of linear size and core dominance $R$. Since our goal is to evaluate core dominance as an orientation indicator, we cannot employ their $\theta_\mathrm{pseudo}$ in our analysis. In Figure \ref{fig:pseudo} we plot $r_\mathrm{geom}$ versus $\theta_\mathrm{pseudo}$ for the Hough-Readhead sample. The two measures are significantly correlated (Spearman $\rho$ = 0.53, p-val = 0.0062), which lends credence to the use of $r_\mathrm{geom}$ as a proxy for orientation in the present work. A useful feature of $r_\mathrm{geom}$ is that it is based on purely geometrical arguments, and is therefore essentially independent of radio source models or theory.


\section{Testing the core dominance measures against geometric rank}
\label{sect:testing}

In Figure~\ref{fig:rgeom} we plot each core dominance parameter against $r_\mathrm{geom}$. The Spearman correlation coefficients and their p-values are listed in Table \ref{table:rgeom}. The correlations of $R$ and $R_V$ with $r_\mathrm{geom}$ are highly significant, but those of $R_\mathrm{Ox}$ and $R_\mathrm{NLR}$ with $r_\mathrm{geom}$ are only marginally so. In Figure~\ref{fig:rgeom} we also fitted a regression line to the data and calculated the variance of the data points about that the line. The variances are also listed in Table~\ref{table:rgeom}. 

We can now address a possible concern that the (unbeamed) optical continuum  emission is not isotropic. For instance, if it is emitted by a disk, then projection would introduce a $\cos\theta$ dependence in the luminosity. Also, patchy obscuration by the edge of the dusty torus might also decrease the continuum luminosity at larger values of $\theta$ (see DiPompeo et al. (2014) for a discussion of these and other effects). That they are not large factors in the present sample is shown by the top two panels in Figure~\ref{fig:rgeom}. Decreasing continuum luminosity with increasing $\theta$ would flatten the slope of the regression line of $R_V$ against $r_\mathrm{geom}$, but it can be seen that the slopes for $R$ and $R_V$ are essentially identical.

Note that the variance for $R_V$ is significantly smaller than that for $R$. Using the Fisher F test, we find that the difference is significant at better than the $1\%$ level. Thus $R_V$ satisfies two of our expectations for a good orientation indicator: it correlates strongly with our geometric proxy for orientation, $r_\mathrm{geom}$, and normalizing by the optical continuum flux introduces significantly less scatter into the correlation than do the other three choices.

To explore this further we conducted the following experiment of randomizing the denominators. We constructed core dominance parameters with a shuffled power measure in the denominator. That is, for each source

\begin{equation}
R_\mathrm{randomized} = \frac{\mathrm{Core\ Luminosity}} {\mathrm{Power\ proxy\ from\ a\ different\ source}}
\end{equation}

\begin{figure*}
\begin{minipage}{0.4\linewidth}
  \includegraphics[width=\linewidth]{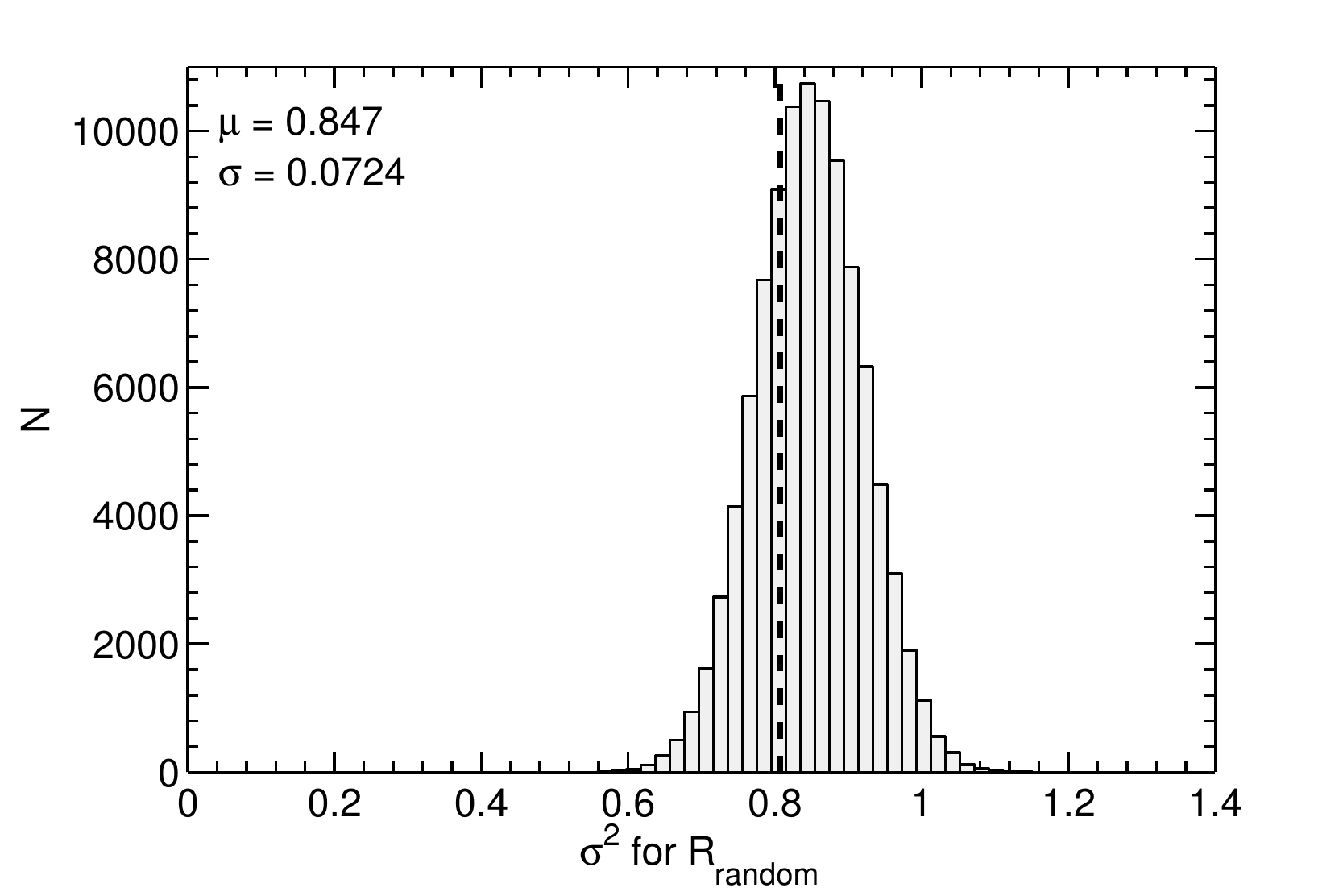}
\end{minipage}
\begin{minipage}{0.4\linewidth}
  \includegraphics[width=\linewidth]{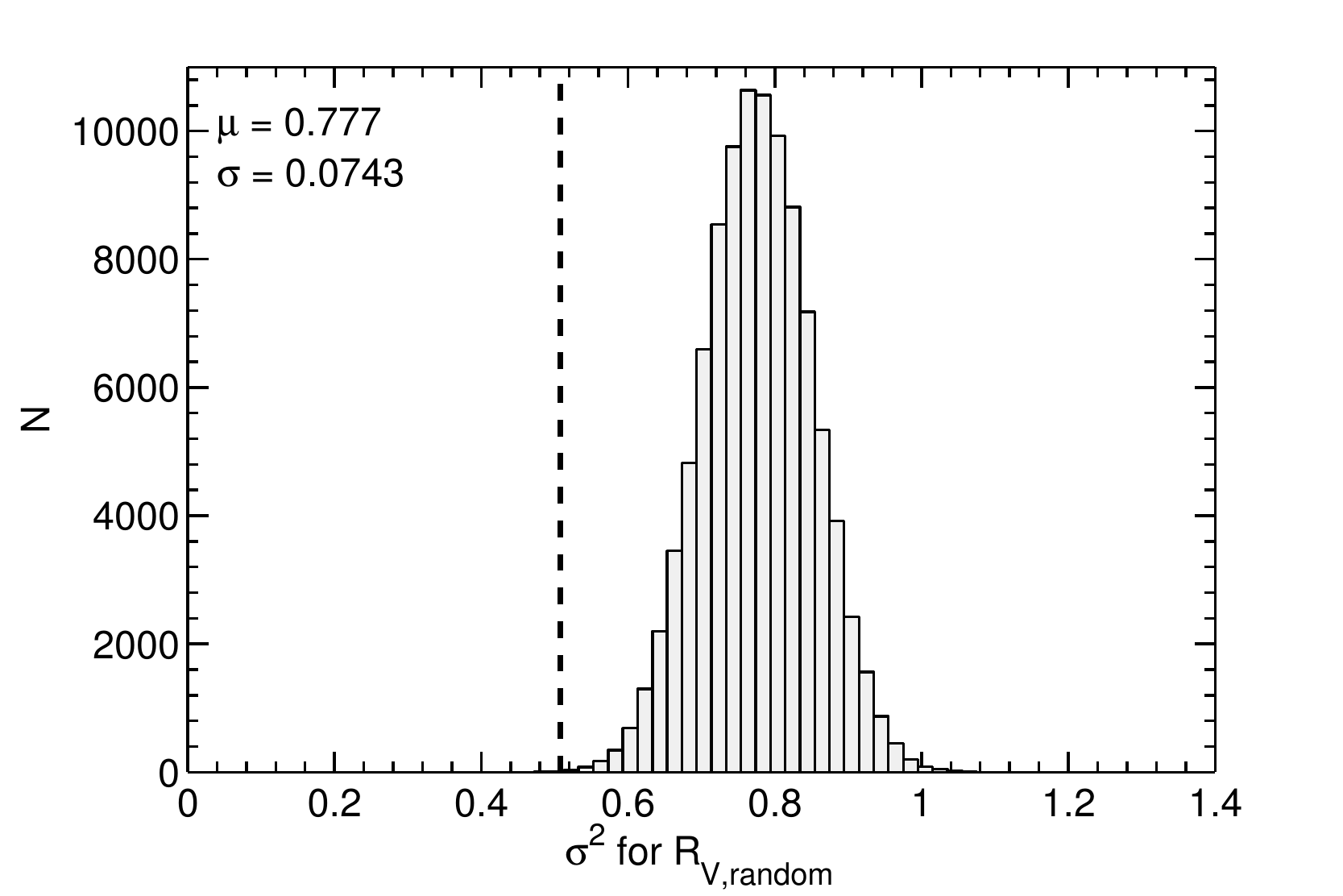}
\end{minipage}
\begin{minipage}{0.4\linewidth}
  \includegraphics[width=\linewidth]{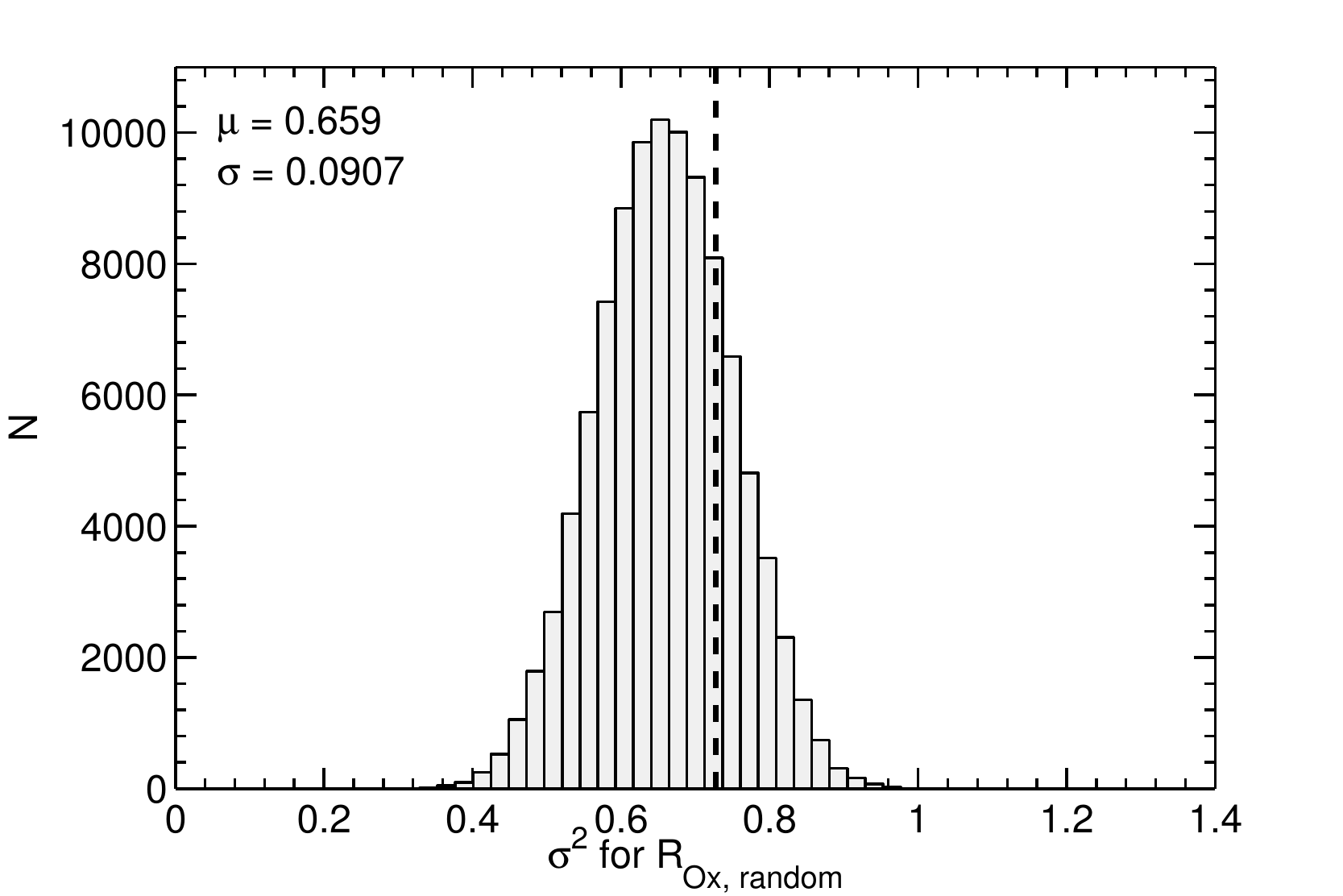}
\end{minipage}
\begin{minipage}{0.4\linewidth}
  \includegraphics[width=\linewidth]{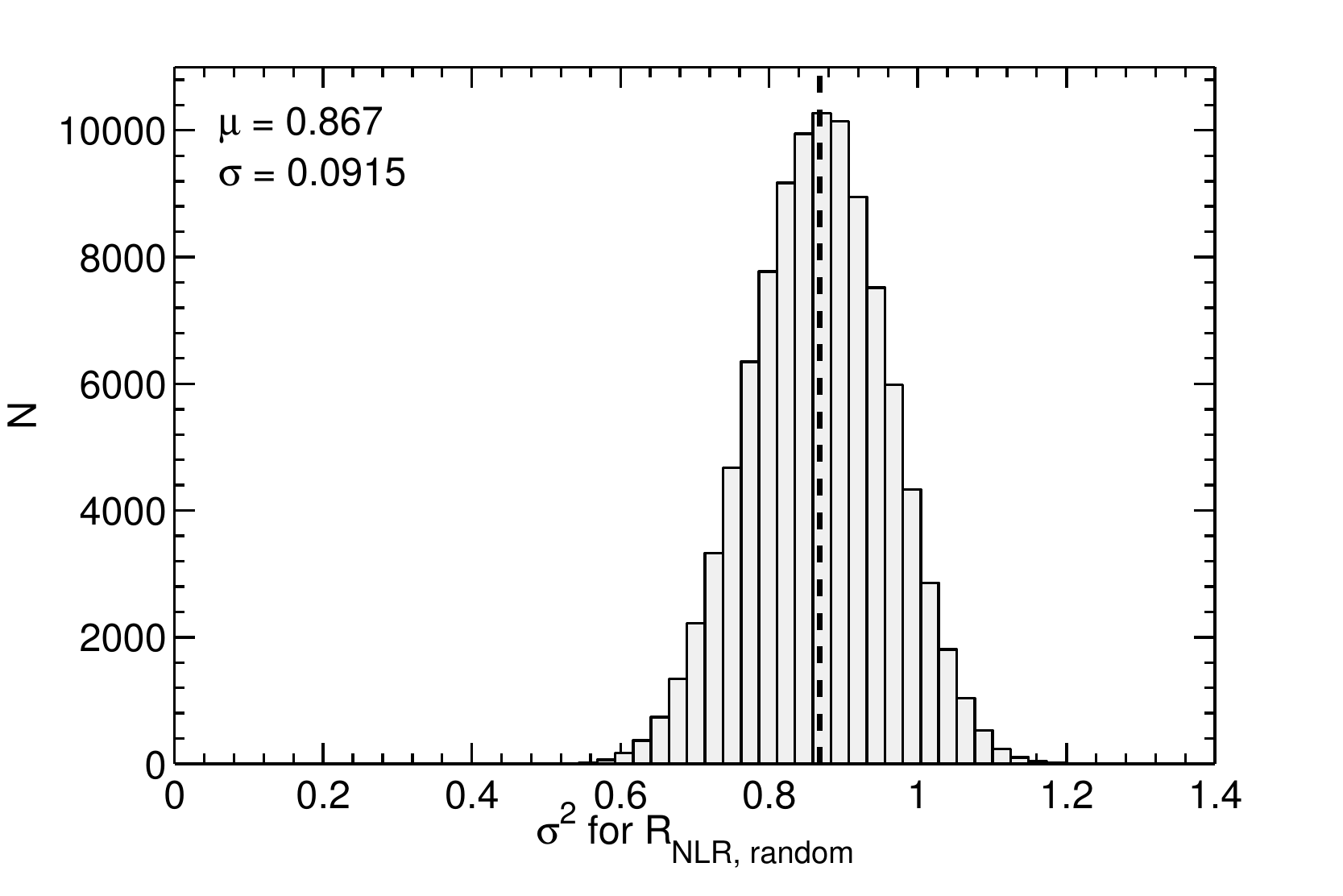}
\end{minipage}
\caption{Histograms of the variance about the best fit line of the randomized core dominance measures, with calculated mean $\mu$ and standard deviation $\sigma$. The vertical lines mark the variance $\sigma^2$ of the non-randomized core dominance (from Table~\ref{table:rgeom}). Only $R_V$ shows a variance that is significantly different from its randomized distribution.}
\label{fig:variance}
\end{figure*}

For each set of shuffled core dominances generated, we calculated the variance about the best fit line (corresponding to Figure~\ref{fig:rgeom}). We repeated this 100,000 times for the randomizations of each of the four possible measures of core dominance, building up a distribution of variances $\sigma^2$. We can now answer to what extent any given proxy for the engine power is an improvement over random scatter. We plot the histograms of the calculated $\sigma^2$ distributions in Figure \ref{fig:variance}. For $R$, $R_\mathrm{Ox}$, and $R_\mathrm{NLR}$, the values listed in Table~\ref{table:rgeom} (where the numerators and denominators belong to the same source) lie within 1$\sigma$ of the mean variance of their randomized counterparts. For $R_V$, however, the value listed in Table~\ref{table:rgeom} lies $3.6\sigma$ below the mean. Under the assumption that this distribution is Gaussian, the probability that the value in Table~\ref{table:rgeom} was pulled from the randomized distribution is 0.0012.  

Figure~\ref{fig:variance} demonstrates strikingly that normalizing a particular source's core radio luminosity by the lobe luminosity, the narrow-line luminosity, or the value of $Q_\mathrm{Ox}$ that pertain to that source, is no better than normalizing by values belonging to any other sources in the sample. In other words, they all fail as useful proxies for the intrinsic engine power of that source. The corresponding core dominance measures still have some merit as orientation indicators, but that stems purely from the orientation dependence of the numerator. The denominators mostly add noise. The exception is $R_V$ where it evidently matters a great deal that the radio core luminosity be normalized by the optical continuum luminosity {\em of the same source}. We therefore conclude that $R_V$ is a demonstrably superior orientation indicator, confirming the original suggestion made by Wills \& Brotherton (1995).

Strictly, what is demonstrated by Table~\ref{table:rgeom} and Figure~\ref{fig:rgeom} is that both $L_{\nu,\mathrm{5GHz\ Core}}$ and $L_{\nu,\mathrm{2500 \AA\ }}$ contain a common variable that cancels when we construct $R_V$. We suggest that this common variable is the underlying intrinsic engine power, which on physical grounds must enter both quantities.

\section{Hough-Readhead Sample}
\label{sect:HRsample}

We have repeated this analysis for the smalller but very well-observed Hough-Readhead (1989) sample of 25 lobe-dominated quasars drawn from the revised 3CR sample (Laing et al. 1983). The criteria for inclusion in the sample are a FR II (Fanaroff \& Riley 1974) radio morphology, and a flux density of the extended emission (presumed to be unbeamed) $> 10$ Jy at 178 MHz. 

We construct the rank $r_\mathrm{geom}$ as before. Here, however, we adopt slightly modified definitions of projected arm length and bending angle, following Hough \& Readhead. We rank on projected total linear size $L$ (adjusted for redshift as before), and curvature $C_b$, the angle produced at the intersection of two lines connecting the points of brightest emission on either side of the source and the core.

We limit our analysis to $R$ and $R_V$. We calculate the scatter about the best fit line and find it to be 0.407 in the case of $R_V$ and 0.442 in the case of $R$. Here, the correlation of $r_\mathrm{geom}$ with $R$ is insigificant ($\rho = -0.22,\ p-value = 0.29$), while the correlation with $R_V$ is significant ($\rho = -0.46,\ p-value = 0.021$). As before, we generate randomized $\sigma^2$ distributions.  The variance of $R_V$ lies 2.3$\sigma$ from the mean of its randomized distribution, while the variance of $R$ lies $1.6\sigma$ away from its own. Fitting the randomized $\sigma^2$ distributions to a Gaussian, we find the probability that $R_V$ is pulled from randomized distribution to be 0.021. For $R$, we find a probability of 0.11.

These results are not as dramatic as for the larger LSS, but they are nonetheless completely consistent with it.

\section{Summary}
\label{sect:conclude}
We have considered four measures of core dominance, each of which normalizes the radio core luminosity by a different observed quantity that purports to serve as a proxy for the intrinsic power of the central engine, and tested them using a lobe-selected sample (LSS) of 126 quasars. We examined the possibility of contamination of the optical continuum flux by beaming. Finally, we discriminated between the competing core dominance parameters by using geometrical information taken from the radio images. Our results are summarized below:

\begin{enumerate}
\item By comparing the mean equivalent widths of the broad-lines H$\beta$, MgII, and CIV for the sources of the LSS with those of the MOJAVE-Mets\"{a}hovi blazar sample and using a simple model for the continuum luminosity, we show that the contribution of beamed synchrotron  radiation from the relativistic jet is not in general a significant component of the optical continuum flux for a sample of sources that has been selected to be as free as possible of orientation bias.
\item We define the rank $r_\mathrm{geom}$ to incorporate both projected arm length and bending angle as geometric proxies for orientation angle. We find that, when plotted against $r_\mathrm{geom}$, $R_V$ produces significantly less scatter about the best fit line than do the other three definitions of core dominance. 
\item By a comparison to core dominance parameters with randomized engine power normalizations, we find that only $R_V$ produces a scatter that significantly differs from its randomized counterpart ($3.6 \sigma$). For this reason, we conclude that the optical continuum luminosity performs better than the competing proxies for engine power at normalizing the central engine power, rendering $R_V$ the superior indicator for orientation.
\item We repeat this analysis for the Hough-Readhead LDQ sample, limiting our analysis to $R$ and $R_V$. The result is consistent with the LSS result---the variance of $R_V$ lies $2.3\sigma$ from the mean of its randomized distribution, while that of $R$ lies only $1.6\sigma$ away.
\end{enumerate}

This is the first of a series of papers that will explore methods to constrain the orientation of FR II radio sources in general and quasars in particular. The range of applications for a reliable measure of quasar orientation is large, as attested to by the (currently) 458 citations to Orr and Browne (1982).

\section*{Acknowledgments}
We thank Mike Brotherton and Bev Wills for reading an earlier draft of this paper and making many helpful suggestions. We also thank Teddy Cheung for providing the original high redshift quasar sample that initiated this and much other work.

JFCW has been supported by NSF grants AST 0607453 and AST 1009261. 

The National Radio Astronomy Observatory is a facility of the National Science Foundation operated under cooperative agreement by Associated Universities, Inc.

This research has made use of the NASA/IPAC Extragalactic Database (NED) which is operated by the Jet Propulsion Laboratory, California Institute of Technology, under contract with the National Aeronautics and Space Administration. 

Funding for SDSS-III has been provided by the Alfred P. Sloan Foundation, the Participating Institutions, the National Science Foundation, and the U.S. Department of Energy Office of Science. The SDSS-III web site is http://www.sdss3.org/.

SDSS-III is managed by the Astrophysical Research Consortium for the Participating Institutions of the SDSS-III Collaboration including the University of Arizona, the Brazilian Participation Group, Brookhaven National Laboratory, Carnegie Mellon University, University of Florida, the French Participation Group, the German Participation Group, Harvard University, the Instituto de Astrofisica de Canarias, the Michigan State/Notre Dame/JINA Participation Group, Johns Hopkins University, Lawrence Berkeley National Laboratory, Max Planck Institute for Astrophysics, Max Planck Institute for Extraterrestrial Physics, New Mexico State University, New York University, Ohio State University, Pennsylvania State University, University of Portsmouth, Princeton University, the Spanish Participation Group, University of Tokyo, University of Utah, Vanderbilt University, University of Virginia, University of Washington, and Yale University.

\label{lastpage}

\end{document}